# Widely tunable, efficient on-chip single photon sources at telecommunication wavelengths

Thang Ba Hoang,[1,*] Johannes Beetz[2], Matthias Lermer[2], Leonardo Midolo[1], Martin Kamp[2], Sven Höfling[2], and Andrea Fiore[1]

[1]*COBRA Research Institute, Eindhoven University of Technology, P.O. Box 513, NL-5600MB Eindhoven, The Netherlands*

[2]*Technische Physik and Wilhelm Conrad Röntgen Research Center for Complex Material Systems, Universität Würzburg, Am Hubland, D-97074 Würzburg, Germany*

We demonstrate tunable on-chip single photon sources using the Stark tuning of single quantum dot (QD) excitonic transitions in short photonic crystal waveguides (PhC WGs). The emission of single QDs can be tuned in real-time by 9 nm with an applied bias voltage less than 2V. Due to a reshaped density of optical modes in the PhC WG, a large coupling efficiency $\beta \geq 65\%$ to the waveguide mode is maintained across a wavelength range of 5 nm. When the QD is resonant with the Fabry-Perot mode of the PhC WG, a strong enhancement of spontaneous emission is observed leading to a maximum coupling efficiency $\beta = 88\%$. These results represent an important step towards the scalable integration of single photon sources in quantum photonic integrated circuits.

*Author to whom correspondence should be addressed. Electronic mail: t.b.hoang@tue.nl

The integration of QDs with PhC structures, such as photonic crystal cavities (PhCCs) and PhC WGs, has attracted tremendous attention in recent years as it may enable the realization of large arrays of efficient single photon sources on a chip. PhC WGs are particularly interesting since the QD-WG coupling does not rely on a sharp spectral resonance, making it more fabrication-tolerant. Indeed, the increased local density of states (LDOS) in the flat part of the dispersion curve of PhC WG modes enhances the QD spontaneous emission rate [1-5], resulting in an optimized coupling efficiency over a relatively broad spectral range. The integration of semiconductor QDs in PhC WGs has been realized and experimentally investigated [2,6-11]. Several works [7,9-11] have examined short GaAs PhC WG structures, where single InAs QD emission lines can be coupled to the WG slow light modes while avoiding disorder-induced localization due to fabrication imperfections [12]. It was also shown that in addition to the enhancement due to the high optical density of states in the slow-light frequency regime, the QD emission is further enhanced due to the resonance with the Fabry-Perot (FP) waveguide cavity modes (resulting from the reflection at the two WG end facets) [4,10]. However, for any practical use in quantum photonic integrated circuits, one needs to be able to tune the single QD emission wavelength precisely in order to produce two-photon interference from distinct QDs [13,14]. Here we report the first electrically tunable single-photon sources in PhC WGs. Due to the broadband nature of the LDOS reshaping in a PhC WG we achieve a high coupling efficiency to the WG mode over a large tuning range of ~ 5 nm, much larger than it would be possible using a fixed-wavelength cavity. Additionally, we observe a strong enhancement of QD emission rate when they cross FP waveguide cavity modes leading to coupling efficiencies $\beta \sim (65-88)\%$. The electrical tunability combined with high efficiencies make these sources very attractive for integration into quantum photonic integrated circuits.

A 1.5 μm thick AlGaAs sacrificial layer and a 320nm thick membrane layer were grown on a (100) undoped GaAs substrate by molecular beam epitaxy. The layer structure is schematically shown in Fig. 1(a). The membrane, which acts as WG layer in the final device, consists of a central InAs QD layer (a few QDs/$\mu m^2$, emitting near 1300 nm [15]) surrounded symmetrically by 5 nm thick GaAs spacers, 20nm thick $Al_{0.33}Ga_{0.67}As$ layers, and 135 nm GaAs layers. The AlGaAs layers act as electronic barriers in order to achieve a large Stark tuning range [16]. During the growth of the membrane, a p-i-n diode was formed by introducing Si-dopants ($2x10^{18}$ $cm^{-3}$) in the bottom 50 nm and Be-dopants ($2x10^{18}$ $cm^{-3}$) in the top 30 nm. In order to allow precise pattern alignment of all subsequent lithography steps, Au markers were first fabricated on the sample. Thereafter, the part of the sample reserved for PhC WGs and p-contacts was defined by electron beam lithography (EBL) and covered with a $BaF_2$/Cr etch mask using electron beam evaporation (EBE) and lift-off. Electron cyclotron reactive ion etching (ECR-RIE) was applied to etch the unprotected area down to the middle of the n-doped layer of the membrane (depth ~ 295 nm). After removal of the etch mask, electrical contacts on both the bottom and top level were patterned using EBL and lift-off process, with a film sequence of 17nm Cr / 33 nm Pt / 333 nm Au deposited by EBE. For the fabrication of the PhC WGs, the sample was covered with 100 nm $SiO_2$ using sputter deposition and the PhC hole pattern was defined (closely to the p-contact) by electron beam lithography using a polymethylmethacrylate resist. The W1 PhC WGs were formed by leaving a row of holes un-patterned. Afterwards, the pattern was transferred from the resist into an underlying $SiO_2$ layer by $CHF_3$/Ar RIE and then into the membrane by ECR-RIE. In order to remove the AlGaAs sacrificial layer below the PhC WGs and the residual SiO$_2$, the sample was immersed in hydrofluoric acid. The WG length was varied between 10 and 25 $\mu m$. Such a WG length is long enough to produce the slow light effect and yet short enough to minimize the effect of disorder due to fabrication imperfections [17, 18]. The PhC lattice spacing *a* was designed to vary around 315 nm in order for the slow-light frequencies of the PhC WGs to be in resonance with the QD ground state emission at low temperature (~1280 nm). Also, in addition to a good current-voltage characteristic of the device, the PhC WG should exhibit good FP modes in the slow-light regime therefore the sample was cleaved before etching. A scanning electron microscopy (SEM) image of a final device is shown in the inset of Fig. 1(b) where a W1 PhC WG is seen lying in between the two p-n electrical contacts.

In the optical measurements, we used a custom designed low-temperature (5K) cryogenic probe station in combination with a micro-photoluminescence ($\mu$-PL) set-up. In this set-up, the sample was mounted on the cold finger of a continuous flow helium cryostat and the QDs in a PhC WG are excited (from the top) by a laser (either 80 MHz pulsed diode $\lambda = 757$ nm laser or cw $\lambda = 780$ nm laser) through a long working distance objective. The PL emission from single QDs was collected from the WG side facet using a lensed single-mode fiber fed through the cryostat body and placed at a distance of approximately 10 $\mu m$ from the WG facet. Both the lensed fiber and the electrical contact probes were mounted on piezoelectric positioners. An illustration of the experimental configuration is shown in Fig. 1(b). After being collected, emitted photons are either dispersed by a f=1m spectrometer and detected by an InGaAs charge-coupled device (CCD) camera or coupled to superconducting single photon detectors

(SSPDs) for time-resolved and anti-bunching measurements [19]. Single QD lines were filtered using the above mentioned spectrometer or by a tunable band-pass filter.

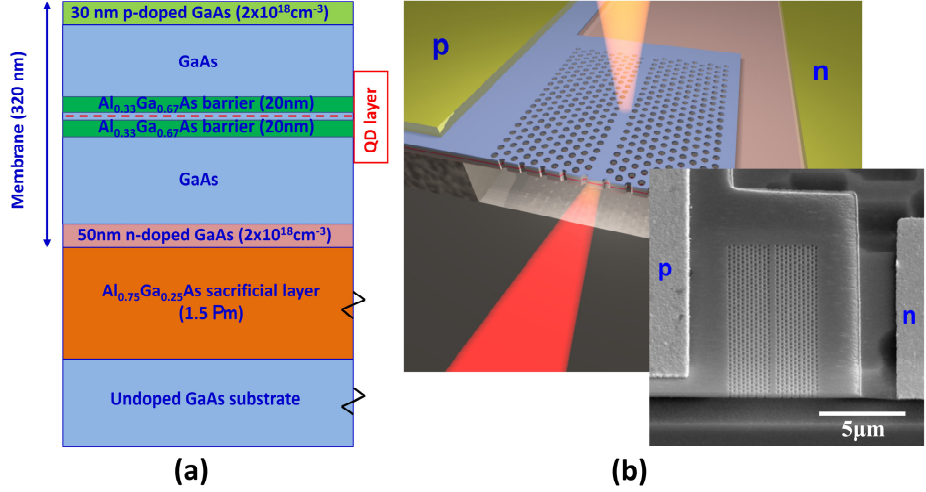

Fig. 1. (a) Layout of the QD sample design (b) Sketch of the device investigated in this study. The inset in the bottom right shows the SEM image of a final device from the sample used for measurements.

We first present in Fig. 2 a $\mu$-PL measurement of several single QD lines in a PhC WG under different DC bias voltages obtained with an optical excitation power ~90 $nW$. The false color image (with red color indicating higher intensity) is constructed from $\mu$-PL spectra taken from the WG side facet while the applied voltage is scanned from 0 to 1.8V in *forward* bias. Indeed, at zero or negative bias the QD emission lines were not observed indicating that the built-in junction field is sufficient to sweep carriers away from the QDs. We therefore use a forward bias to reduce and control the electric field across the QDs. From the current-voltage characteristic of the WG device, shown as an inset in Fig. 2, one can clearly see that within the range from 0 to 1.8 V the current through the device is negligible. The FP mode profile of the PhC WG device (the white spectrum at the bottom of the image) can be easily observed if one increases the laser excitation power to around $100\,\mu W$. The strong PL enhancement of single QDs at the FP frequencies is due to the combination of the slow group velocity and the spatial confinement created by the reflection at the WG end facets, as previously reported [10]. It is clearly seen in Fig. 2 that the single narrow QD lines are tuned to the blue when the bias voltage is increased (i.e. field across QD decreased) and a tuning range of as much as 9 nm (7 meV) is observed for several QD lines. It is also very important to note that when changing the applied bias voltage the PhC WG FP modes do not tune within the measurement resolution. As a result of these tuning characteristics (i.e QDs vs FP modes) the emission of single QD lines can be sequentially tuned into resonance with several FP modes by varying the bias voltage. When a QD line crosses a FP mode, its intensity is strongly enhanced, due to the increased LDOS at the frequencies of the FP modes.

The quantum confined Stark tuning of InAs QDs (both single dot lines and ensemble) resulting from an applied vertical field and the existence of a permanent dipole have been reported before [20-24]. Consistently with previous reports [20,21], the blue shifting of single dot lines observed in our experiments indicates an inverted electron-hole alignment (i.e hole wave-function located above the electron wave-function along the growth direction). However, it is not possible in our case to accurately estimate the QD permanent dipole moment and the polarizability as there is a significant voltage drop in the contacts (as seen in the turn-on voltage of the current-voltage curve in the inset of Fig. 2), so that the internal field cannot be determined precisely.

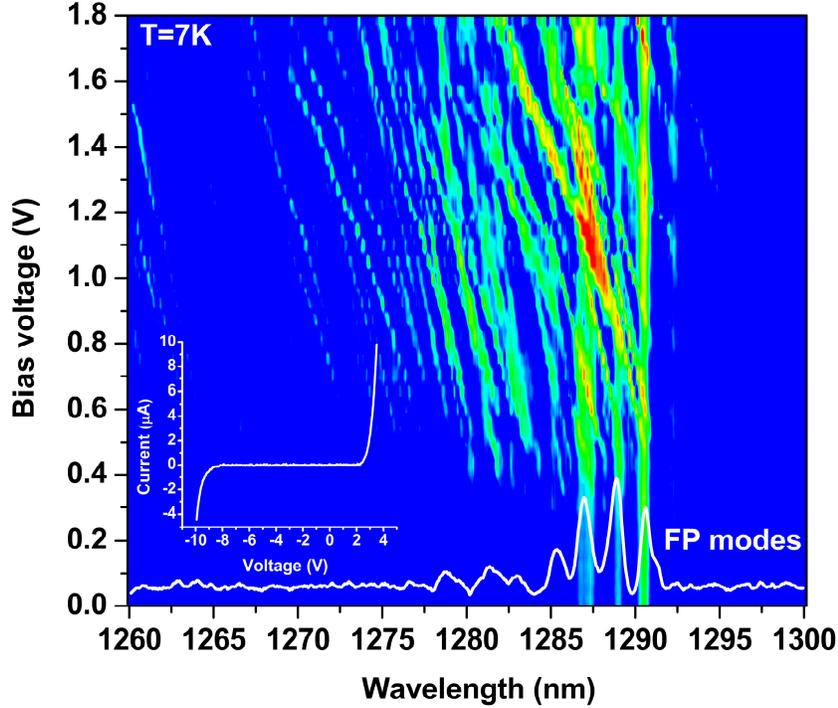

Fig. 2. False color image (red indicates higher intensity) shows the QD PL emission wavelength (measured from the WG side facet) as a function of the applied forward bias voltages. The white spectrum at the bottom of the image shows the FP mode profile if the PhC WG is excited with high laser excitation power (~100 $\mu W$ ). It is clearly seen that the QD lines are blue shifted when the bias voltage increases while the FP resonance do not shift. The inset shows the current-voltage characteristic of this particular device.

In order to explore the tuning dependent dynamics of the QDs, we measured the decay times of single QD lines as a function of the bias voltages. In this experiment, a pulse diode laser ($\lambda$=757 nm, pulse width = 75 ps, repetition rate = 80 MHz) with very low excitation power (P=20 $nW$) is used so that the FP mode emission is negligible when the QDs are out of resonance, ensuring that emission from single QD lines is measured and not the cavity-enhanced background [25]. In Fig. 3 we show the decay times of several single QD lines measured from the WG side facet while the bias voltage is changed from 0.5 to 1.8 V. The decay times are deduced from fitting the time-resolved decay curves (e.g red curve in Fig. 4) including the convolution with the measured instrument response function. For an easy comparison we also show in Fig. 3 the PL spectrum detected at high excitation power (P=50 $\mu W$) where several FP modes are clearly visible. We consistently observed that when a single QD is tuned, the decay time of the dot decreases each time it is resonant with a FP mode. The QD decay times decrease from around 1.7 ns, when the QD emission is away from the resonance, down to ~ 0.6 ns as the emission wavelength approaches the band-edge similarly to our previous observations in similar short PhC WG and to the theoretical predictions [4,10]. The decreased decay time is a strong evidence of the increasing of the LDOS in the slow light regime as compared to the rest of the dispersion curve.

Two important factors determining the emission characteristics of single QDs in a PhC WG are the Purcell enhancement factor $F_P$ and the coupling efficiency $\beta$-factor between the QD and the WG modes. The Purcell factor describes how much the spontaneous emission rate of QDs in the PhC WG is enhanced compared to the QDs in bulk. Here we determined that the Purcell factor $F_P \sim 2$ when the dot is tuned into resonant with the FP mode at 1290.5 nm (near the band edge) and taking into account the measured lifetime of 1.4 ns of these QDs in bulk GaAs (green curve in Fig. 4) and the decay rate into leaky modes as discussed below. This limited value of $F_P$ is quite typical of PhC WGs [7,8,10,11], and reflects the larger mode volume and lower Q-factor (ranging from 1500 at low wavelength to nearly 3000 at the band edge in the present study) as compared to the PhC cavities, besides spatial mismatch between QD and WG modes.

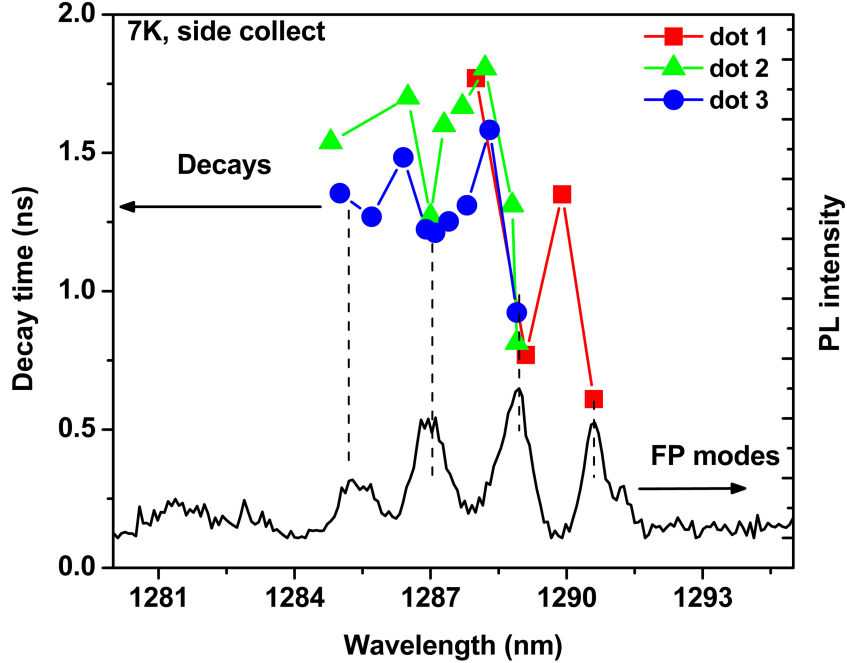

Fig. 3. Decay times (left axis) of several single QDs measured from the WG side facet at different bias voltages (0.5-1.8 V). The PL spectrum (black, right axis) detected at high excitation power showing clearly several FP modes. The decay times of all single QDs decreased each time they are in resonance with the FP modes. Solid and dashed lines are guides to the eyes.

We also performed the decay time measurements of single QDs when they do not couple to the PhC WG and QDs in the PhC membrane (outside of the WG region) by collecting these dots from top through the cryostat window in a different, top-collection μ-PL setup. For example, in Fig. 4 we show the decay curves of three dots measured in different situations: a dot (dot 1 in Fig. 3) which is tuned in resonance with a FP mode at 1290.5 nm (collected from the WG side facet), a dot in bulk and a (uncoupled) dot in the PhC outside of the WG (collected from top). The decay times of these three dots are 0.61 ns, 1.4 ns and 6.1 ns, respectively. We mention here that due to the large dispersion of the QD positions in a PhC membrane, their decay times vary significantly [6]. In our experiments we observed a typical range of 3-6.1 ns and an average of 5 ns can be made. This value is comparable with previous reports [6,8]. The coupling efficiency $\beta$ of the QD emission into the PhC WG slow light mode, can be calculated as $\beta = 1 - \frac{\tau_{WG}}{\tau_{PhC}}$. Here $\tau_{WG}$ is the measured decay time of QDs coupled to the WG mode and $\tau_{PhC}$ is the average decay time of uncoupled QDs from the same device. The time $\tau_{PhC}$ quantifies the effect of emission rate into leaky modes and nonradiative recombination. For the data presented in Fig. 3, we can estimate a coupling efficiency $\beta > 65\%$ across the entire investigated tuning range of ~5 nm for all three QDs. The highest $\beta$ reaches 88% for QD1 when it is tuned in resonance with the FP mode at 1290.5 nm. The high $\beta$ value indicates an efficient coupling between the QD emission and the WG slow light FP modes and this is crucially important for channeling photons from QDs into WGs for integrated quantum photonic applications. We emphasize that such a spectrally-broad coupling efficiency, demonstrated here for the first time, is a unique feature of 1D PhC WG systems, related to the broad nature of the LDOS reshaping in the WG. In QD-PhC cavity systems, in contrast, tuning of both the QD and cavity would be needed to realize such a widely tunable single-photon-source.

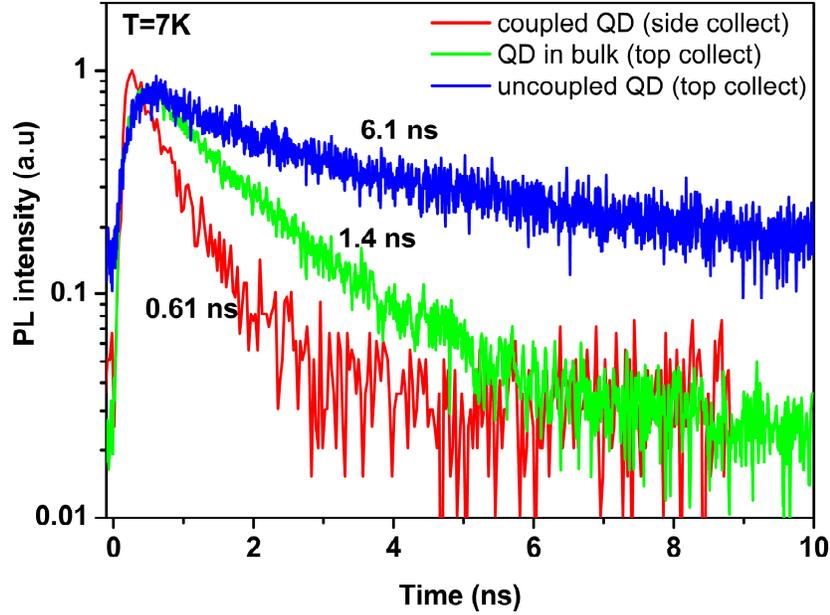

Fig. 4. PL decays of single QDs at different configurations: coupled to a WG FP mode (measured from side facet), in bulk or uncoupled (measured from top).

Finally, in order to demonstrate the nonclassical nature of the emission of single QDs in PhC WG configuration, we performed an antibunching measurement. The second-order correlation function $g^{(2)}(\tau)$ of the same excitonic line at 1287.2 nm presented in Fig. 2 (at the applied bias of 1.1 V when the QD is in resonance with a FP mode and at lower laser excitation power P=30 nW) was measured by coupling the QD emission from the WG side facet into a single-mode fiber, and fiber-based band-pass filter. The single QD line was then sent to a fiber-based Hanbury-Brown and Twiss interferometer using two SSPDs [19] and electronic signals were analyzed by a PicoHarp 300 module. A clear dip appeared at time $\tau = 0$ (see Fig. 5) showing antibunching and thus proving the quantum nature of the single QD emission. The fit (red curve), using $g^{(2)}(\tau) = 1 - Ae^{-\tau/t}$ with A and t are the fitting parameters, to the measured data returned a value $g^{(2)}(0) = 0.36$ and decay time of 0.93 ns. This decay time value is close to the decay times presented in Fig. 3 when the QDs are resonant with the third FP mode (~ 1287.2 nm). The residual $g^{(2)}(0)$ is likely related to a residual multiphoton background emission which is not completely filtered due to limited spectral width of the tunable filter (0.8 nm bandwidth).

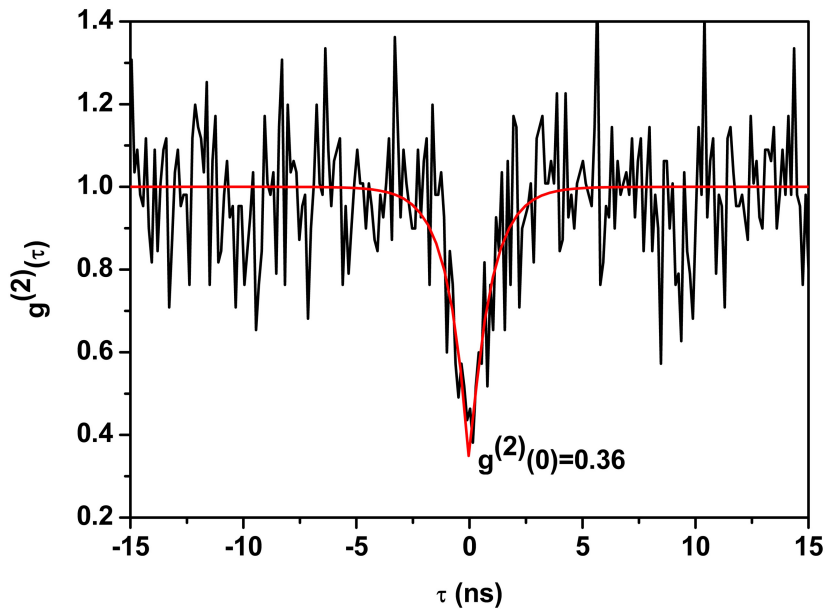

Fig. 5. Autocorrelation measurement on a single dot line (at 1287.5 nm, 1.1 V of the data set presented in figure 2) showing nonclassical light emission. Red curve is the fit to the normalized measured data (black).

Single QD lines are electrically tuned by up to 9 nm using the Stark effect and can be brought into resonance with several FP modes of the short PhC WG. From both integrated and time-resolved PL experiments we observe a strong increase of the spontaneous emission rate at the FP mode resonances. From measured decay times of single QDs we derived a coupling efficiency $\beta \geq 65\%$ over a broad spectral range of 5 nm, and a maximum $\beta = 88\%$ at resonance. The result of our study is a critical step on road toward the realization of a quantum photonic integrated circuit with on-chip tunable single photon emitters.

**Acknowledgments** We acknowledge financial support from the European Commission within the FP7 project QUANTIP (project n. 244026) and from the Dutch Technology Foundation STW, applied science division of NWO and the Technology Program of the Ministry of Economic Affairs.